\documentclass[twocolumn,showpacs,preprintnumbers,prc,superscriptaddress,nofootinbib]{revtex4}
\usepackage{epsfig}
\usepackage{graphicx}
\usepackage{amsmath,amssymb,amsfonts}
\usepackage{array}
\usepackage{url}
\usepackage{hyperref}
\usepackage{lineno}
\usepackage{xspace}
\usepackage[usenames,dvipsnames]{color}

\graphicspath{{./pics/}}

\newcommand{\com}[1]       {\relax}
\newcommand{\PbPb}         {\mbox{Pb--Pb}}

\newcommand{\pPb}          {\mbox{p--Pb}}

\newcommand{\dAu}          {\mbox{d--Au}}

\newcommand{\pt}           {\ensuremath{p_{\mathrm{T}}}{ }}

\newcommand{\snn}          {\ensuremath{\sqrt{s_{\mathrm{NN}}}}}

\newcommand{\Npart}        {\ensuremath{N_\mathrm{part}}}
\newcommand{\avNpart}      {\ensuremath{\langle \Npart \rangle}}

\newcommand{\apart}        {\ensuremath{\alpha_\mathrm{part}}}
\newcommand{\aspec}        {\ensuremath{\alpha_\mathrm{spec}}}

\newcommand{\abs}[1]       {\ensuremath{\left|#1\right|}}
\newcommand{\avg}[1]       {\ensuremath{\left\langle#1\right\rangle}}

\newcommand{\dd}           {\ensuremath{\mathrm{d}}}
\newcommand{\dNdy}         {\ensuremath{\frac{\dd N}{\dd y}}}
\newcommand{\Tab}[1]       {Tab.~\ref{#1}}
\newcommand{\Fig}[1]       {Fig.~\ref{#1}}
\newcommand{\Figure}[1]    {Figure~\ref{#1}}

\newcommand{\Section}[1]   {Section~\ref{#1}}

\newcommand{\Eq}[1]        {Eq.~\ref{#1}}

\newcommand{\Ref}[1]       {Ref.~\cite{#1}}
\newcommand{\Refs}[1]      {Refs.~\cite{#1}}

\begin{document}

\title{Effects of Longitudinal Asymmetry in Heavy-Ion Collisions}
\author{Rashmi Raniwala}
\email{Rashmi.Raniwala@cern.ch}
\affiliation{Physics Department, University of Rajasthan, Jaipur, India}
\author{Sudhir Raniwala}
\email{raniwalasudhir@gmail.com}
\affiliation{Physics Department, University of Rajasthan, Jaipur, India}
\author{Constantin Loizides}
\email{Constantin.Loizides@cern.ch}
\affiliation{Lawrence Berkeley National Laboratory, Berkeley, California, USA}
\date{\today}

\begin{abstract}
In collisions of identical nuclei at a given impact parameter, the number of nucleons participating in the overlap region of each nucleus can be unequal due to nuclear density fluctuations.  
The asymmetry due to the unequal number of participating nucleons, referred to as longitudinal asymmetry, causes a shift in the center of mass rapidity of the participant zone.
The information of the event asymmetry allows us to isolate and study the effect of longitudinal asymmetry on rapidity distribution of final state particles.
In a Monte Carlo Glauber model the average rapidity-shift is found to be almost linearly related to the asymmetry.
Using toy models, as well as Monte Carlo data for Pb--Pb collisions at 2.76 TeV generated with HIJING, two different versions of AMPT and DPMJET models, 
we demonstrate that the effect of asymmetry on final state rapidity distribution can be quantitatively related to the average rapidity shift via a third-order polynomial with a dominantly linear term.
The coefficients of the polynomial are proportional to the rapidity shift with the dependence being sensitive to the details of the rapidity distribution.
Experimental estimates of the spectator asymmetry through the measurement of spectator nucleons in a Zero Degree Calorimeter may hence be used to further constrain the initial conditions in ultra-relativistic heavy-ion collisions.
\end{abstract}
\maketitle


\section{Introduction}
\label{sec:intro}
In collisions of heavy ions the geometrically overlapping region created by interacting nucleons from each nucleus is called the participant zone. 
Even at fixed impact parameter, the number of participating nucleons from each nucleus fluctuates around the mean due to fluctuations in the positions of the nucleons around the mean nuclear density profile.  
Event-by-event, the participant zone therefore has a net non-zero
momentum in the nucleon-nucleon centre-of-mass (CM) frame, and hence
its rapidity is shifted with respect to the CM frame, and is denoted
by $y_0$~\cite{Vovchenko:2013viu}.

Experimental data and simulated data from different event generators
show that the total produced particle multiplicity, measured over a wide phase space
region, scales approximately with the number of participants~\cite{Alver:2010ck,Abbas:2013bpa}. 
The observed $\Npart$ scaling indicates that the number of
participants or wounded nucleons is a relevant parameter affecting the 
production and distribution of produced particles even at LHC energies.
Fluctuations of the fireball shape in the longitudinal direction are
expected to create nontrivial rapidity correlations, as explicitly
demonstrated using the wounded nucleon model~\cite{Bzdak:2012tp}.
The different components of the fluctuating fireball shape suggested
in ~\cite{Bzdak:2012tp}, have been recently extracted from the
measured two-particle rapidity correlations ~\cite{Jia:2015jga}. 

The distribution of charged particles averaged over a large number of
events in collisions of identical nuclei is observed to be symmetric
about the rapidity of the nucleon-nucleon CM frame. 
The observed forward backward asymmetry in collisions of 
non-identical nuclei \dAu~\cite{Steinberg:2007fg} and \pPb~\cite{Martinez-Garcia:2014ada}
has been argued to be due to shift of the rapidity of the participant zone. 
Unequal number of participants in a collision of two identical nuclei
produces a shift of the participant zone. This shift can be observed
by measuring asymmetry in the energy of the Zero Degree Calorimeters on
either side of the interaction vertex, even though its estimate in central
collisions is marred by large relative
fluctuations; the measurement of shift may facilitate to
separate effect of fluctuations on various observables~\cite{Csernai:2012mh}.
Preliminary results of the ALICE collaboration show a difference
between pseudorapidity distribution of charged particles in \PbPb\
collision events of
different asymmetries, as estimated from the measurement of spectator
neutrons in the neutron Zero Degree
Calorimeters~\cite{Raniwala:2015vfo}. 
Simulations based on a fluid dynamical framework demonstrate 
the effect of longitudinal fluctuations on the azimuthal anisotropy coefficients and their
rapidity dependence;  a significant decrease in the values of $v_1(y)$, and a wide
plateau like behaviour for $v_2(y)$, both near midrapidity, has been
estimated due to the event-by-event fluctuations affecting the rapidity of the
participant zone, the latter being a conserved quantity ~\cite{Csernai:2011gg}.
The similar transverse-momentum dependence of the rapidity-even directed flow and the corresponding estimate from two-particle correlations at mid-rapidity indicate a weak correlation between fluctuating participant and spectator symmetry planes and suggest the possibility of using the spectator nucleons to further determine and constrain the effect of initial conditions~\cite{Abelev:2013cva}.
This possibility has recently been further explored by model studies
using AMPT~\cite{Bairathi:2015sga}. 

It has been argued that the
vorticity arising due to initial state angular momentum may survive the evolution process in a low viscosity 
state of quark-gluon plasma and may manifest in the rapidity
dependence of directed flow, $v_1(y)$, however its
observability is affected by initial state fluctuations necessitating
the requirement to determine the event-by-event centre of mass~\cite{Csernai:2013bqa}.
The need for determination of the event-by-event centre of mass
rapidity is also highlighted by the possible observation of
$\Lambda$-polarisation in the CM frame, which may confirm attainment of local thermodynamical
equilibrium and persistence of vorticity until freezeout of the
expanding matter~\cite{Becattini:2013vja,Xie:2017upb}.
The observation of $\Lambda$-polarisation at the Relativistic Heavy
Ion Collider requires that the models describing the evolution of a heavy-ion
collisions incorporate the effect of large vorticity, thereby
providing a complete characterisation of the system necessary to
understand the dynamics of quarks and gluons in extreme conditions~\cite{STAR:2017ckg}.

In the present work, we investigate the possible effect of the net~(non-zero) momentum of the participant zone on experimentally measurable distributions of produced particles by exploring possible correlations between the participant asymmetry and the distribution of particles in the kinematic phase space. 
Events of the same net-momentum can be selected by classifying events
on the basis of measured asymmetry in spectators for any centrality. 
The rapidity distribution of events of any asymmetry class is studied
relative to corresponding distributions of another asymmetry class. 
The method has the advantage that most experimental uncertainties and 
corrections affecting the single particle distributions are
cancelled. 
All variables used to develop this analysis can be estimated experimentally.

The paper is organized as follows. 
The rapidity-shift of the participant zone due to asymmetry of the
event is discussed in \Section{sec:rapshift}. 
The effect of a rapidity-shift is estimated using a toy model on a
Gaussian rapidity distribution and is discussed in \Section{sec:constrap}.
\Section{sec:varrap} discusses the effect on various charged particle
rapidity distributions for variable rapidity-shifts as calculated
using Glauber model.
The results of the present work are summarized in \Section{sec:conc}.

\section{Rapidity Shift of Participant Zone}
\label{sec:rapshift}
If the number of nucleons participating from the two colliding nuclei is $A$ and $B$, respectively, then the participant zone has a net momentum in the nucleon-nucleon CM frame.
The net momentum corresponds to a shift in the rapidity of the participant zone, which can be approximated as
\begin{equation}
\label{y0def}
y_0 \cong \frac{1}{2} \ln \frac {A}{B}.
\end{equation}
Assuming each of the $A$~($B$) nucleons has a fixed momentum $p$~($-p$), \Eq{y0def} is obtained using the sum of four-momentum vectors $(0,0,Ap,AE)$ and $(0,0,-Bp,BE)$, with $E^2=m_0^2 + p^2$, and neglecting $m_0\ll p$.
Since at the LHC in the TeV scale, $m_0/p<10^{-6}$, we replace the  '$\cong$' sign by the equality sign hitherto.

Defining the asymmetry of participants for each event as $\apart = \frac{A-B}{A+B}$, the rapidity-shift $y_0$ can be written as 
\begin{equation}
\label{y0alpha}
y_0 = \frac{1}{2} \ln \frac{1+\apart}{1-\apart}\,,
\end{equation}
and has a unique correspondence with $\apart$. 
For small $\apart$, the shift follows $y_0 \cong \apart$. 
The unequal number of nucleons in the participant zone imply unequal number of spectators of the two colliding nuclei, $N-A$ and $N-B$, respectively, where $N$ is the total number of nucleons in each nucleus.
The spectator asymmetry $\aspec = \frac {(N-A) - (N-B)}{(N-A) + (N-B)} = \frac{B-A}{2N-(A+B)}$ is related to the participant asymmetry via $\aspec = -\apart\frac{A+B}{2N -(A+B)}$.
Finally, the rapidity shift $y_0$ is related to the spectator asymmetry as 
\begin{equation} 
\label{y0spect}
y_0 = \frac{1}{2} \ln \frac{(A+B)(1+\alpha_{spec})- 2N \aspec}{(A+B)(1-\aspec)+ 2N\aspec}\,
\end{equation}
which is accessible to experiment.
Unlike the unique correspondence between $\apart$ and $y_0$, the presence of the $(A+B)$ term in \Eq{y0spect} leads to a distribution of $y_0$ for a given value of $\aspec$, even at fixed impact parameter or centrality.

\begin{table}[t!]
\begin{center}
\begin{tabular}{l||c|c|c|c}
\hline
Centrality &$b_{\rm min}$~(fm) & $b_{\rm max}$~(fm) & $\avNpart$ & $\abs{y_{0}}$ \\ 
 \hline\hline
  0--5\% &0.0  &3.61 & 383.9 & 0.0144\\ 
 5--10\% &3.61 & 5.12&330.0 & 0.0263\\   
10--15\% &5.12 & 6.27& 280.2&  0.0352\\   
15--20\% & 6.27& 7.24& 236.5& 0.0431\\  
20--25\% & 7.24 &8.09 & 198.5& 0.0512\\  
25--30\% & 8.09&8.86 & 165.5& 0.0589\\  
30--35\% & 8.86& 9.57& 136.5& 0.0678\\   
35--40\% & 9.57& 10.23& 110.2& 0.0777\\  
40--45\% &10.23 & 10.85& 88.6&  0.0887\\   
45--50\% & 10.85& 11.43& 70.2& 0.1012\\  
50--55\% & 11.43 & 11.99& 54.6& 0.1155\\  
55--60\% & 11.99& 12.52& 41.6& 0.1326\\  
60--65\% & 12.52&13.03 & 31.0& 0.1528\\   
65--70\% & 13.03& 13.52& 22.5& 0.1764\\  
 \hline
\end{tabular}
\caption{Centrality classes defined by impact parameter as well as corresponding $\avNpart$ and $\avg{\abs{y_0}}$ values.}
\label{table:ZDCMeanRMS}
\end{center}
\end{table}

\begin{figure*}[t!]
\begin{center}
\includegraphics [width=1.0\textwidth]{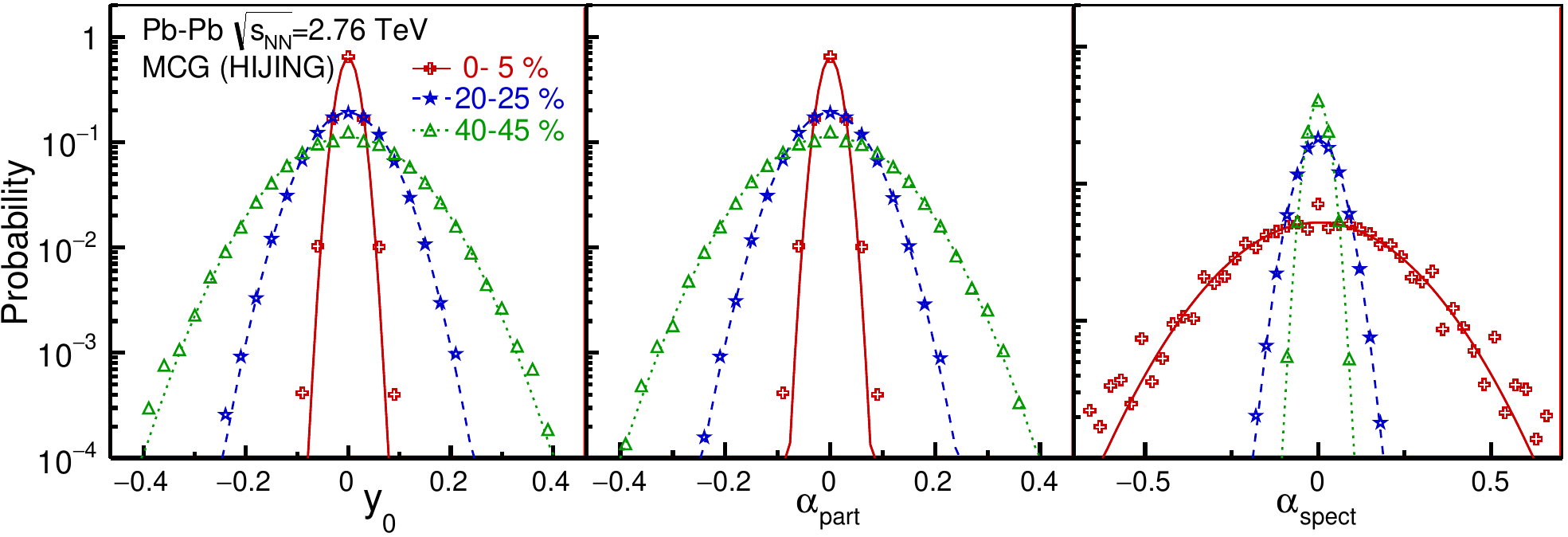}
\caption{The distributions of the participant-zone rapidity shift $y_0$~(left panel), of the participant asymmetry $\apart$~(middle panel) and of the spectator asymmetry $\aspec$ for 0--5\%, 20--25\% and 40--45\% \PbPb\ centrality classes calculated with the HIJING MCG.}
\label{fig:AsymDis}
\end{center}
\end{figure*}

%

The rapidity-shift $y_0$~(\Eq{y0def}) as well as its dependence on $\apart$~(\Eq{y0alpha}) and $\aspec$~(\Eq{y0spect}) can be calculated within a Monte Carlo Glauber~(MCG) framework~\cite{Miller:2007ri}, as implemented in \Refs{Alver:2008aq,Loizides:2014vua} or in \Ref{Wang:1991hta}.
In the present work, we have generated 1.2 million minimum bias events of \PbPb\ collisions at $\snn = 2.76$ TeV using HIJING~(v1.383)~\cite{Wang:1991hta} with default settings. 
The generated impact parameter~($b$) is used to define the event centrality. 
Limits on $b$ used for 5\%-wide centrality intervals with corresponding $\avNpart$ are provided in \Tab{table:ZDCMeanRMS}.  
The mean number of participants provide an estimate of the order of magnitude of the rapidity-shift. 
Assuming that the number of nucleons from each of the two nuclei fluctuate by their root mean square while retaining the total number to be equal to $\Npart$, the resulting value of rapidity shift for the most central class of events would be $\approx0.05$. 
In practice, events in any centrality class will have a distribution peaked at zero. 
The distributions of $y_0$, $\apart$ and $\aspec$ calculated with the HIJING MCG are shown in \Fig{fig:AsymDis} for three centrality classes along with a Gaussian fit for each.  
The width of the $y_0$ distribution increases with decreasing centrality, i.e.\ for larger impact parameters the relative fluctuations increase since the number of participants decreases. 
Events can be classified according to their rapidity-shift: $y_0<0$ are events of negative asymmetry {\it(-asym)} and $y_0>0$ are events of positive asymmetry {\it(+asym)}.  
For each centrality class, the mean value of $\abs{y_0}$ is also reported in \Tab{table:ZDCMeanRMS}. 
The increase of $\abs{y_0}$ with decreasing collision centrality is in agreement with results obtained earlier~\cite{Vovchenko:2013viu}.
The $\apart$ distributions are nearly identical to the $y_0$
distributions, while the $\aspec$ distributions are different.
The widths of the $\aspec$ distributions increase with increasing centrality, i.e. the relative fluctuations increase with decreasing number of spectator nucleons.

\begin{figure}[t!]
\includegraphics[trim=0.0cm 0.0cm 0.0cm -1.5cm,clip, width=\linewidth]{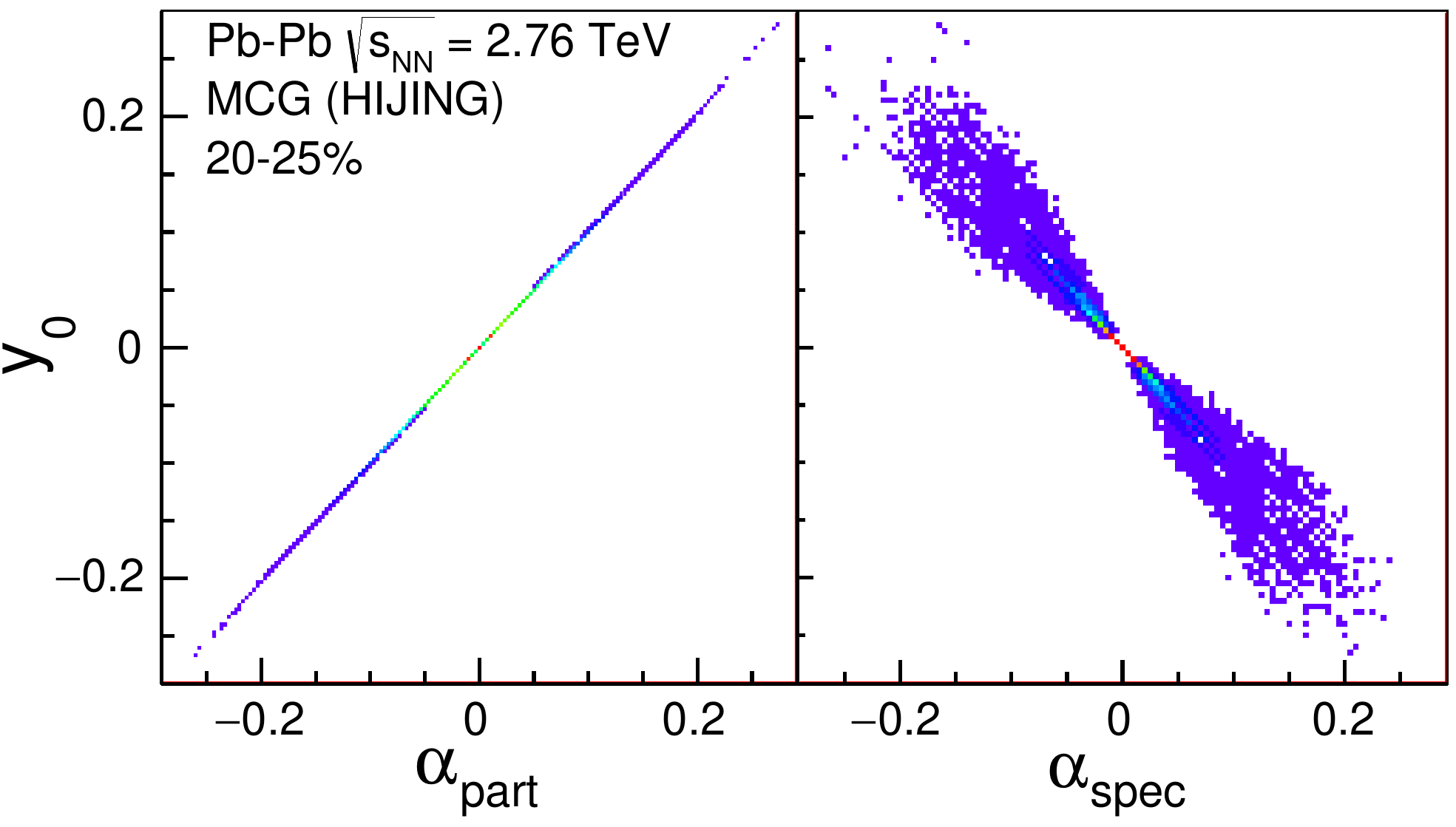}
\caption{Event-by-event distributions of $y_0$ versus $\apart$~(left panel) and $y_0$ versus $\aspec$ for the 20--25\% \PbPb\ centrality class calculated with the HIJING MCG.}
\label{fig:yvsa}
\end{figure}

\begin{figure}
\includegraphics[trim=0.0cm 0.1cm 0.0cm -0.3cm,clip, width=0.85\linewidth]{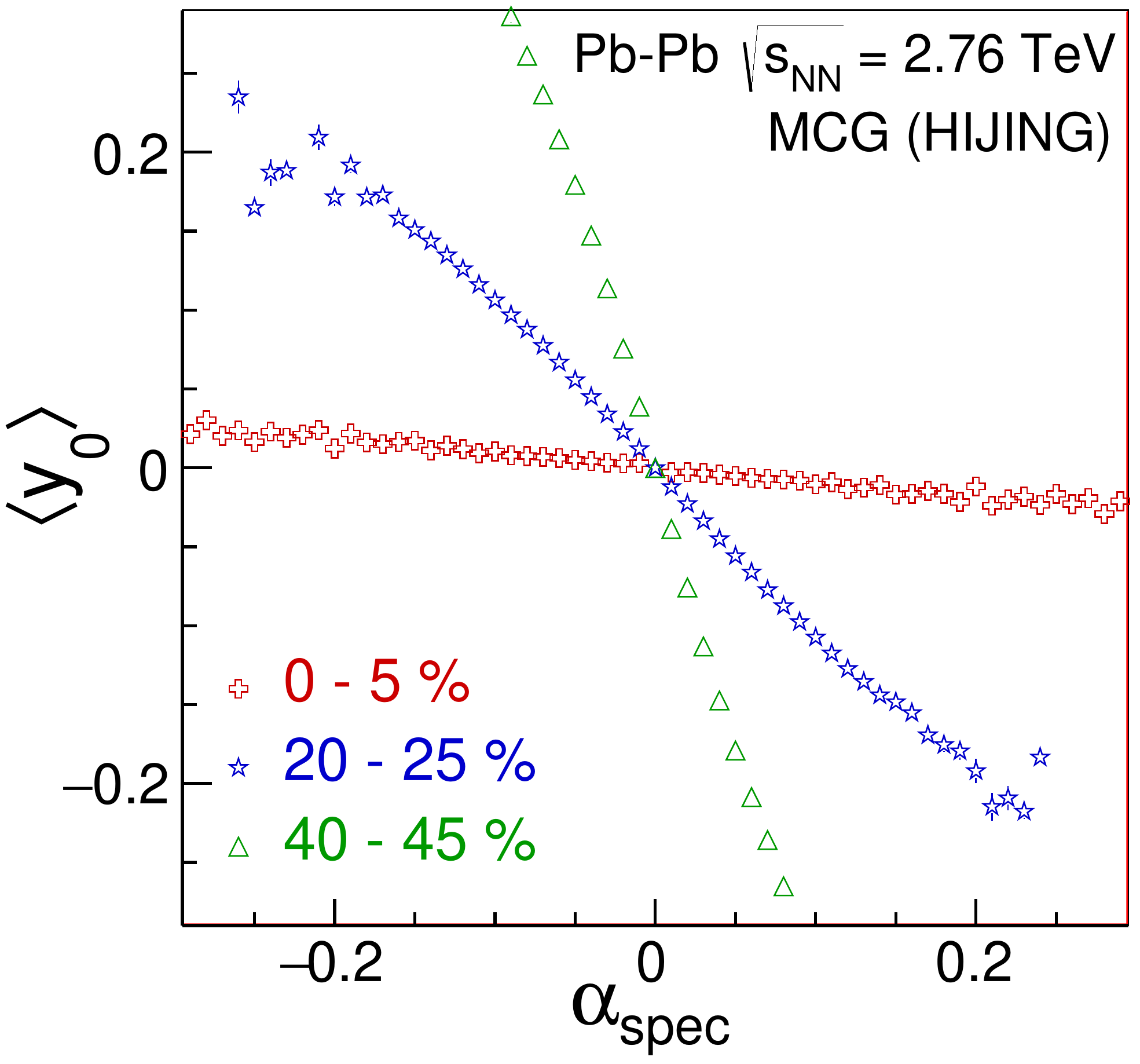}
\caption{The mean participant-zone rapidity shift $\avg{y_0}$ versus $\aspec$ for 0--5\%, 20--25\% and 40--45\% \PbPb\ centrality classes calculated with the HIJING MCG.}
\label{fig:absy0vsa}
\end{figure}

\Figure{fig:yvsa} displays event-by-event distributions of $y_0$ versus $\apart$ and $\aspec$, respectively, obtained for the 20--25\% \PbPb\ centrality class with the HIJING MCG.
The unique correspondence between $y_0$ and $\apart$ is illustrated in the left panel of \Fig{fig:yvsa}. 
The lack of a unique relation between $y_0$ and $\aspec$ due to the presence of the $(A+B)$ term in \Eq{y0spect} leads to a distribution of $y_0$ for a given value of $\aspec$, even at fixed impact parameter or centrality,
as illustrated in the right panel of \Fig{fig:yvsa}.
It can be regarded as the response matrix to obtain the values of $\avg{y_0}$ for a given range of $\aspec$.   
The mean values $\avg{y_0}$ as a function of the $\aspec$ asymmetry are shown in \Fig{fig:absy0vsa} for three different centralities for 0--5\%, 20--25\% and 40--45\% \PbPb\ centrality classes calculated with the HIJING MCG.

If the experiments could measure the number of nucleons in the participant zone, $A$ and $B$, the participant-zone rapidity shift $y_0$ could be determined for each collision.
However, neither $A$ and $B$, nor $\apart$ is directly amenable to experimental measurement. 
The asymmetry $\aspec$ can be estimated by measuring the number of spectator nucleons through their energy deposited in the zero degree calorimeters on either side of the interaction vertex in collider experiments~\cite{Csernai:2012mh}.
Using unfolding methods and the estimated values of  $\aspec$,  one
can obtain an estimate of $\avg{y_0}$, e.g. by using the response
matrix in the right panel of \Fig{fig:yvsa}. Almost all
  estimates of $y_0$ based on Glauber like models will show similar results even if there
  are differences in details. The ALICE experiment has determined a response
  matrix using information on the number of neutrons in
  the spectator and also using the energy deposited in the Zero Degree
Calorimeter, where a Glauber Monte Carlo model
has been tuned to reproduce the energy distributions in neutron Zero
Degree Calorimeters~\cite{Acharya:2017zkw}

\section{Constant Rapidity Shift and Gaussian Charged Particle Rapidity Distribution}
\label{sec:constrap}
The measured rapidity distribution of charged particles produced in collisions of identical nuclei can be described by distributions which are symmetric about the CM rapidity~\cite{Alver:2010ck,Abbas:2013bpa}.
A Gaussian form is amongst the more common distributions used to describe data. 
We assume that the particles produced in asymmetric collisions of identical nuclei are also distributed symmetrically in the CM frame of the participant zone. 
Considering that the rapidity of the participant zone is shifted by $y_0$ from the rapidity of the nucleon-nucleon CM system, a symmetric distribution in the participant zone will appear as a shifted distribution in the nucleon-nucleon CM frame, which is also the laboratory frame for most collider experiments. 
For fixed $y_0$, the rapidity distributions of produced particles can
be written as a Gaussian distribution of width $\sigma$.
\begin{equation}
\dNdy = N_0 \exp (- \frac{(y-y_0)^2}{2\sigma^2})\,.
\end{equation}
For symmetric collisions $y_0=0$, while for longitudinally asymmetric collisions $y_0$ is finite.
The positive and negative values of $y_0$ correspond to the net
momentum of the participant zone in the positive and negative
direction, respectively, causing positive and negative
participant~($\apart$) asymmetries, respectively. 
Taking the ratio of single particle rapidity distributions of different asymmetry classes will eliminate the uncertainties arising due to experimental corrections and fluctuations affecting event-by-event distribution.
The ratio of the rapidity distribution of particles in collisions with positive asymmetry to the distribution in collisions of negative asymmetry yields
\begin{equation}
\begin{split}
\frac{\left(\dNdy\right)_{\rm +asym}}{\left(\dNdy\right)_{\rm -asym}} & = \exp(\frac{4yy_0}{\sigma^2})\\
                                        & = \sum_{n=0}^{\infty} c_n^{\rm g}(y_0, \sigma) y^n \,,
\end{split}
\label{RatioEqn}
\end{equation}
where $c_n^{\rm g}=(4y_0/\sigma^2)^n/n!$ are the coefficients of the Taylor expansion of the exponential function, and the superscript {\it g} of the coefficients stands for the Gaussian shape of the parent rapidity distribution. 
\begin{figure}[h!]
\begin{center}
\includegraphics[width=0.46\textwidth]{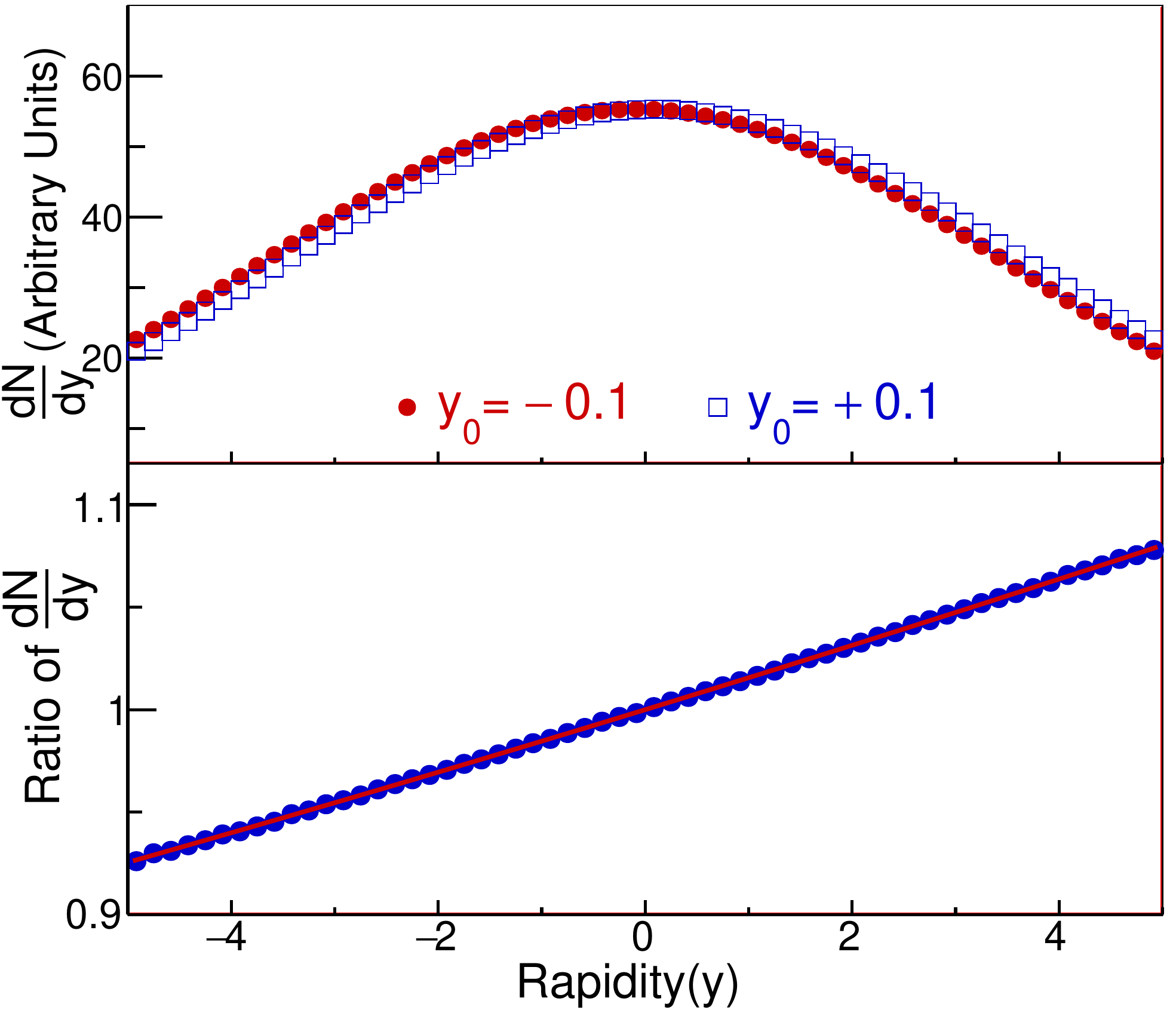}
\caption{Gaussian $\dNdy$ distributions for $\sigma=3.6$ shifted by $y_0\pm0.1$ obtained by the toy model (top panel). Ratio of the $\dNdy$ distributions fitted to a third-order polynomial~(bottom panel). The coefficients are $0.015$, $1.1\times10^{-4}$ and $2.2\times10^{-7}$, respectively with $\chi^2/{\rm dof}=125/117$.}
\label{fig:Gauss}
\end{center}
\end{figure} 
The coefficients depend upon the parameters of the parent rapidity
distribution and on the rapidity shift $y_0$.
These parameters are effectively fixed by selecting events on the basis of centrality and asymmetry. 
For typical values of $y_0$ equal to $0.1$ and $\sigma$ equal to $3.6$, the ratio $\frac{c_2^g}{c_1^g} \sim 0.015$ and the ratio $\frac{c_3^g}{c_1^g} \sim 0.00015$ with subsequent terms having negligible contribution to the values of the function describing the ratio of the two rapidity distributions.  
Hence, for a Gaussian rapidity distribution, where the relation between the shift of the participant-zone rapidity and the coefficients $c_n$ are analytically known, the dominant contribution can be expected for the linear term.
The linear term is related to the rapidity-shift via $y_0=\frac{c_1^g\sigma^2}{4}$.

\begin{figure}[th!]
\begin{center}
\includegraphics[width=0.46\textwidth]{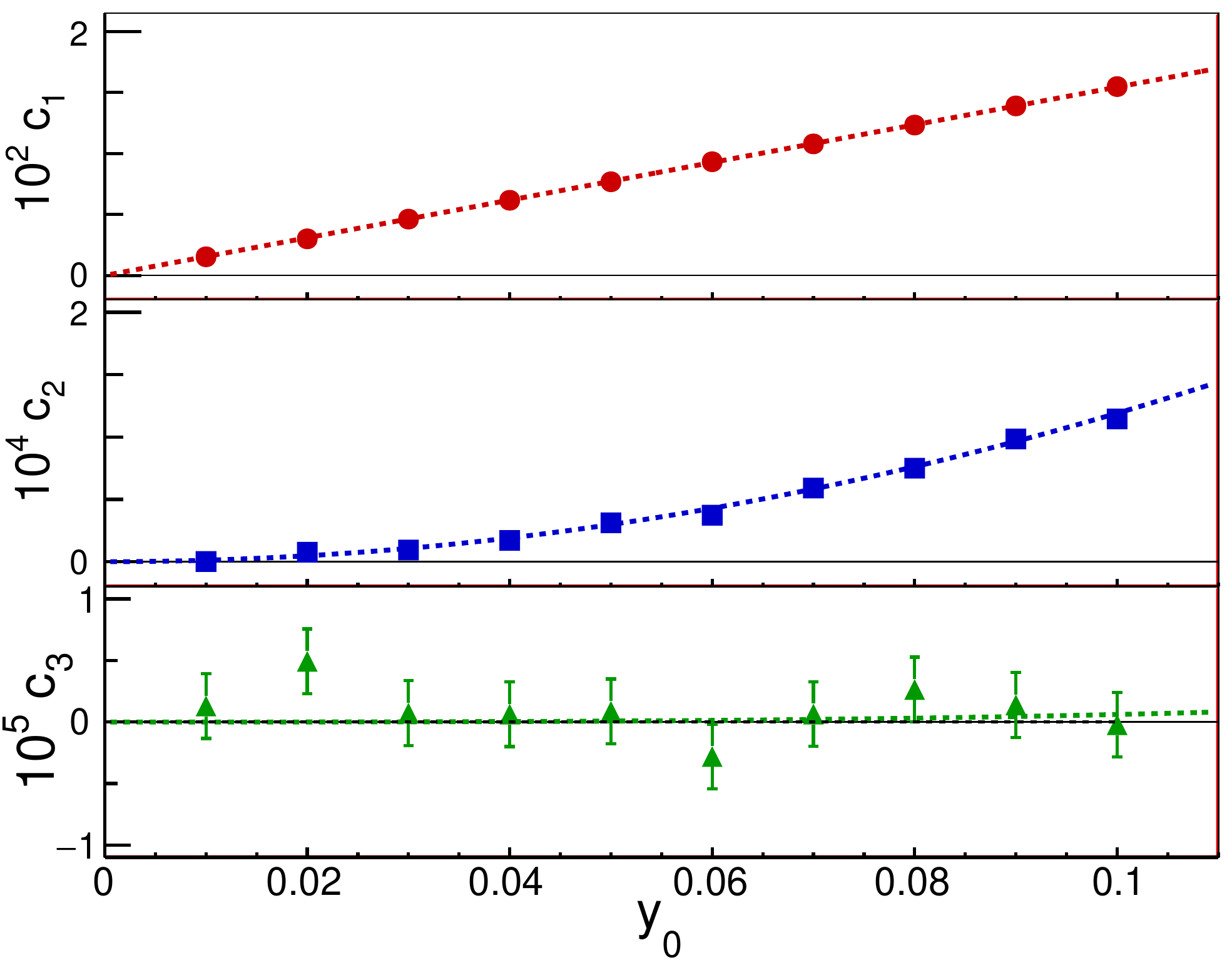}
\caption{The dependences of $c^{\rm g}_1$, $c^{\rm g}_2$ and $c^{\rm g}_3$ on $y_0$ for fixed $\sigma=3.6$ obtained by the toy model. Dashed lines correspond to analytical functions of the coefficients from \Eq{RatioEqn}.}
\label{fig:GaussCn}
\end{center}
\end{figure}
\begin{figure*}[t!]
\includegraphics[width=0.8\textwidth]{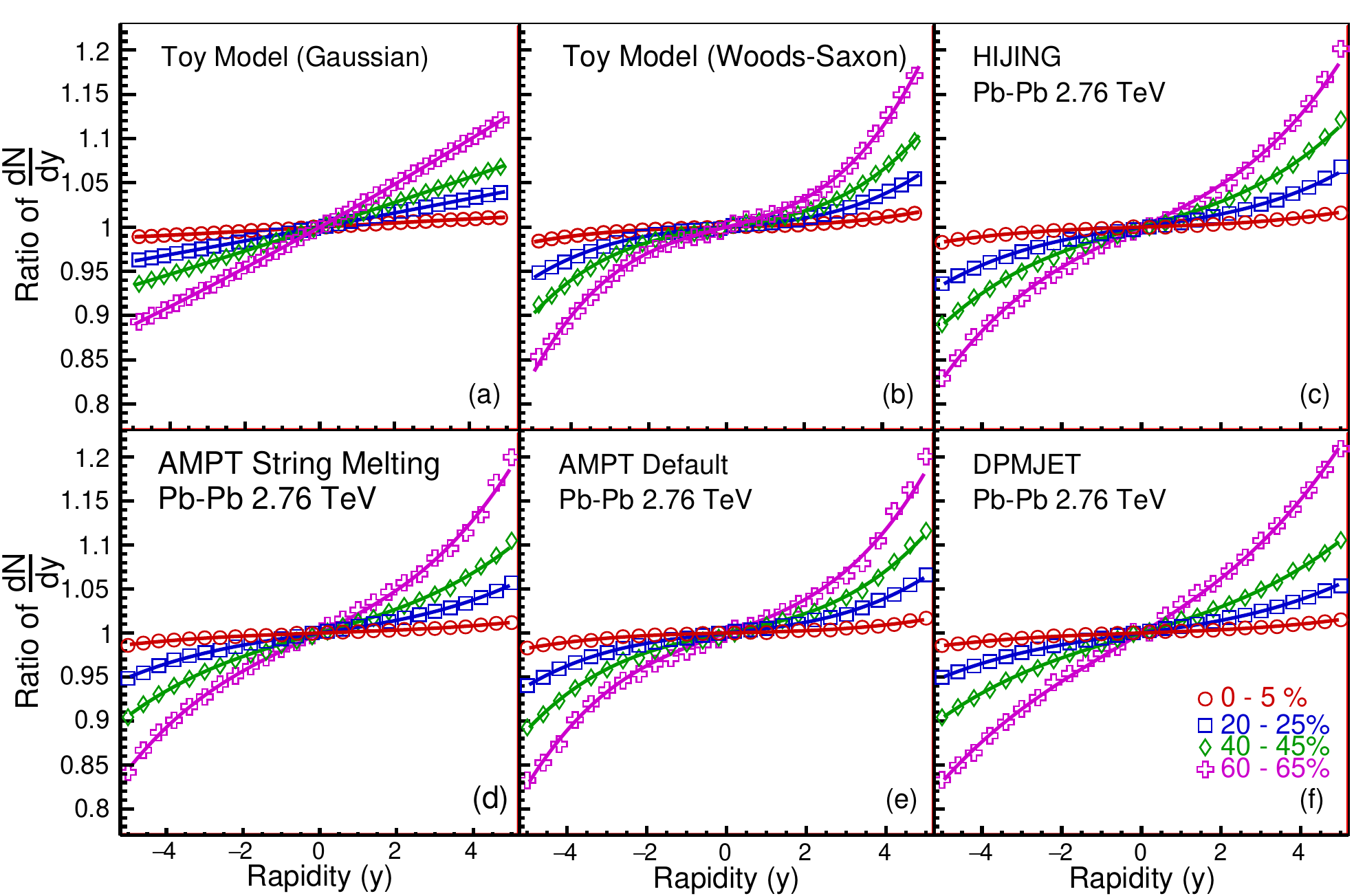}
\caption[onecolumngrid]{The ratio of $\dNdy$ distributions for events with positive~($y_0>0$) and with negative~($y_0<0$) asymmetry. The
  different panels are for rapidity distributions corresponding to  (a) parametrised Gaussian form (b)  parametrised Woods-Saxon form (c) HIJING (d) AMPT (Default)
  (e) AMPT (String Melting) and (f) DPMJET (default). The ratios, along with fits to third order polynomials, are shown for four centrality classes.}
\label{fig:RapRatios}
\end{figure*} 

These results have been validated using a toy model simulation by generating Gaussian rapidity distributions which are shifted by a constant magnitude. 
The top panel of \Fig{fig:Gauss} shows two rapidity distributions for $\sigma=3.6$ shifted by $y_0\pm0.1$. 
The unshifted distribution would obviously lie in between the two distributions and is not drawn.
The bottom panel of \Fig{fig:Gauss} shows the ratio of $\dNdy$ distributions for events with $y_0=+0.1$ to those with $y_0=-0.1$, and the ratio is fitted to a third-order polynomial.
The coefficient of the linear term is dominant with the other coefficients smaller by 2 and 4 orders of magnitude, respectively, as expected from \Eq{RatioEqn}. 
The same calculation is repeated for different values of the shift $y_0$ to obtain the coefficients of the third-order polynomial fit as function of $y_0$, as shown in \Fig{fig:GaussCn}.
The dependence of the coefficients on the rapidity-shift known from \Eq{RatioEqn} is indicated with dashed lines, and agrees very well with the numerical calculation.
The figure demonstrates that the dominant contribution to the $\dNdy$ ratio arises from the first coefficient, which is linearly related to the rapidity shift $y_0$.

\section{Variable Shift and Various Charged Particle Rapidity Distributions}
\label{sec:varrap}
In the previous section we discussed the relation of coefficients of
the third-order polynomial fit to the ratio of $\dNdy$ distributions,
assuming that the $\dNdy$ distributions are Gaussian in nature, and
that all events have the same value of $y_0$.
In the following, the values of $y_0$ for different events are chosen according to the distribution of $y_0$ from the HIJING MCG, as shown for a few centrality classes in the left panel of \Fig{fig:AsymDis}.
As before, the rapidity distribution of particles $\dNdy$ are generated using a toy simulation taking into account event-by-event the rapidity-shift $y_0$. 

In addition to a Gaussian shape for the rapidity distribution, inspired by the possibility
of a double Woods-Saxon distirbution~\cite{Eskola:2002qz}, or a Woods-Saxon-like
distribution~\cite{Alver:2010ck}, we also consider that the
rapidities of produced particles can be described by a Woods-Saxon
distribution.
\begin{equation}
\dNdy = N_0\frac {1}{1+\exp\left(\frac{|(y-y_0)|-a}{c}\right)}
\end{equation}
where $y_0=0$ for symmetric events, positive for events with positive asymmetry, and negative for events with negative asymmetry. 
Taking the ratio of the $\dNdy$ for events of opposite asymmetry, and making a Taylor expansion about $y=0$ yields a polynomial in $y$, which can be written as  
\begin{equation}
\frac{\left(\dNdy\right)_{\rm +asym}}{\left(\dNdy\right)_{\rm -asym}} = \sum_{n=0}^{\infty} c_n^{\rm ws}(y_0, a, c) y^n \,.
\end{equation}
The coefficients $c_n^{\rm ws}$ depend on the shift in rapidity and the parameters of the Woods-Saxon distribution in a non-trivial way. 
For a given set of parameters, however, the dependence on the rapidity-shift $y_0$ can be computed numerically.

We investigate the systematic effect of the rapidity-shift on the
coefficients charecterising the ratio of Gaussian and Woods-Saxon
rapidity distributions for different asymmetry classes.
Using the parameterized form of experimental $\eta$ and $\pt$
distributions in conjunction with the relative yield of pions, kaons
and protons~\cite{Abbas:2013bpa,Abelev:2013vea}, toy-model simulations
provide the rapidity distribution. 
This resulting $\dNdy$ distribution is fitted once to a Gaussian form
and then to a Woods-Saxon form to obtain the values of the parameters.
Using these values of the parameters and the $y_0$ distribution
obtained from HIJING events (\Fig{fig:AsymDis}), the rapidity
distributions are obtained separately for positive and negative values
of $y_0$ corresponding to different centrality classes. The ratio of the two
distributions is fitted to a third order polynomial as shown in
\Fig{fig:RapRatios} for four centrality classes. In each case, 
the $\chi^2/dof$ are the smallest in the most central class and are
about 0.64 and 0.83 for Gaussian and Woods-Saxon rapidity distribution. 

To investigate the possible contribution originating from the dynamics of the particle
production mechanism, or in general, of any final state effects, the charged-particle rapidity distribution
obtained from some commonly used event generators have also been used.
We have generated about 1.2 million events of HIJING and about 1 million events each of
both versions of AMPT (Default and String Melting)~\cite{Lin:2004en}, and of DPMJET~\cite{Roesler:2000he} 
for \PbPb\ collisions at $\snn = 2.76$ TeV. 
The
rapidity shift $y_0$ is determined for each generated event from the
number of participating nucleons from each of the two colliding
nuclei. The average rapidity distributions corresponding to positive and negative
values of $y_0$ are obtained. The ratio of these two rapidity
distributions is fitted to a third order polynomial to obtain the
values of the coefficients. The ratios and the fits  are shown in
\Fig{fig:RapRatios}, along with the results for toy model simulations for
parametrised rapidity distributions. 
The differences in the parent rapidity distributions manifest
themselves in the ratios and the values of the coefficients in the
polynomial.

\begin{figure}
\begin{center}
\includegraphics[width=0.47\textwidth]{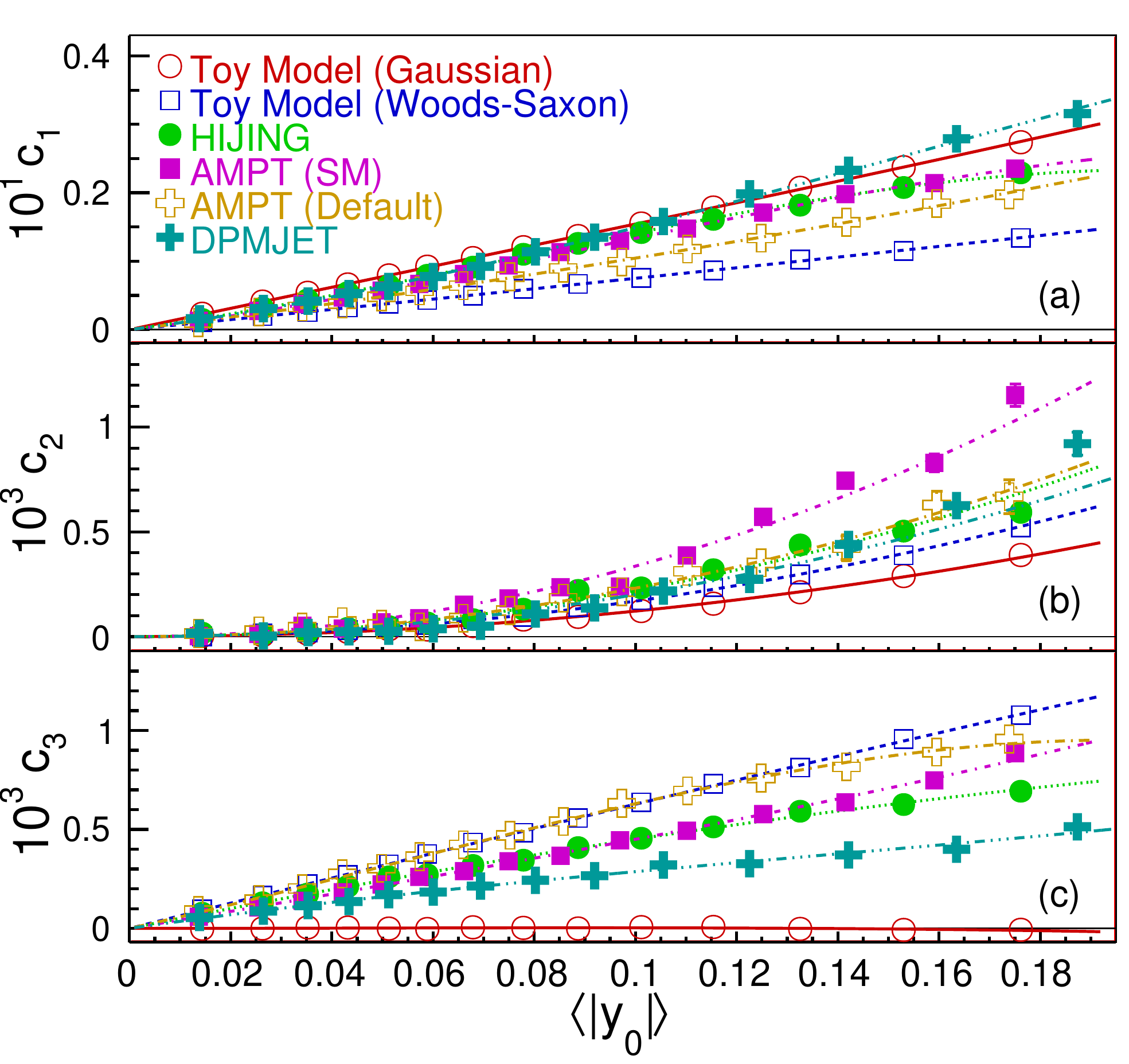}
\caption{The coefficients $c_1$, $c_2$, and $c_3$ of the third-order
  polynomial fit to the ratio of $\dNdy$ distributions versus
  $\avg{\abs{y_0}}$ for the six forms of rapidity distributions. Polynomial fits are to guide the eye. The
  coefficients $c_n$ demonstrate a dependence
  on $\avg{\abs{y_0}}$. 
The rapidity distributions are obtained in 14 centrality intervals up to 70\% centrality with corresponding $\avg{\abs{y_0}}$ as given in \Tab{table:ZDCMeanRMS}. 
}
\label{fig:coeff}
\end{center}
\end{figure}

For all rapidity distributions, the coefficients for the quadratic and cubic
terms in the ploynomial fitting of the ratio are much smaller than those of the linear term.
The dependence of the first three coefficients $c_n$ on the mean
rapidity shift $\avg{\abs{y_0}}$ is shown in \Fig{fig:coeff}. 
The six different rapidity distributions yield different dependence of
the coefficients on $\avg{\abs{y_0}}$, indicating a possibility of
determining the details of the rapidity distribution from the
knowledge of the behaviour of the coefficients. 

\section{Summary}
\label{sec:conc}
In collisions of identical nuclei at a given impact parameter, the
number of nucleons participating in the overlap region of each
nucleus, estimated with a Monte Carlo Glauber model, can be unequal
due to nuclear density fluctuations~(\Fig{fig:AsymDis}).  
The asymmetry due to the unequal number of participating nucleons causes a rapidity-shift of the participant zone~(\Fig{fig:yvsa}), which may be experimentally accessible by measuring the energy of the spectator nucleons, as was also argued in \cite{Csernai:2012mh}.
The average rapidity-shift has been found to be almost linearly related to the asymmetry~(\Fig{fig:absy0vsa}).
The effect of the small rapidity-shift in the rapidity distributions 
was estimated by taking the ratio of the distributions of events of positive and
negative asymmetries. The success of the method was demonstrated by
using a toy model to illustrate that such a ratio can effectively be described by a
third-order polynomial~(\Fig{fig:Gauss}), where the coefficients are
related to the rapidity-shift and are shown for Gaussian rapidity
distributions for constant rapidity-shift~(\Fig{fig:GaussCn}).
The effect on the ratio of rapidity distributions for positive and
negative asymmetry has been systematically studied
for Gaussian and Woods-Saxon particle rapidity distributions, using a
toy model simulation and taking the distribution of $y_0$ from
\Fig{fig:AsymDis}. The ratios and the polynomial fits are shown in
\Fig{fig:RapRatios} (a) and (b). 
The possible effect of dynamics of
particle production is investigated using the ratio of rapidity
distributions for positive and negative asymmetries from different
event generators. The ratios and the fitted polynomials for rapidity
distributions obtained from HIJING,
AMPT and DPMJET are shown in \Fig{fig:RapRatios} (c) to (f) for four
centrality classes. 
The ratios are  sensitive to the detailed shape of the parent rapidity
distribution~(\Fig{fig:RapRatios}), and can be quantitatively
described by a third-order polynomial with a dominantly linear
term~(\Fig{fig:coeff}). The relation between coefficients and the
rapidity shift confirm that the effect of initial state longitudinal
asymmetry survives through the particle production process and the subsequent
evolution to the observed final state.
Experimentally, estimates of the longitudinal asymmetry via
measurements of the spectator asymmetry can be used to systematically
investigate the influence of the longitudinal asymmetry on various
observables, and hence may further constrain the initial conditions in
ultra-relativistic heavy ion collisions. 
Recent results from the ALICE
collaboration confirm that the longitudinal fluctuations affect the
pseudorapidity distributions; the effect finds a simple explanation
in terms of the rapidity shift of the participant zone, and shows a
sensitivity to the shape of the rapidity
distribution~\cite{Acharya:2017zkw}. As indicated in
Sec.~\ref{sec:intro}, determination of an event-by-event CM may be
necessary for a complete description of the evolution of heavy-ion
collisions, to separate the effects of initial state fluctuations  from
the dynamical evolution. Further attempts to devise methods for the determination
of the CM of the participant zone in each event,
using experimentally measurable quantities, is under investigation.

\vspace{1.0cm}
\section{Acknowledgements}
We thank Jurgen Schukraft for fruitful discussions. R.\ Raniwala and S.\ Raniwala acknowledge the financial support of the Department of Science and Technology of the Government of India.
The work of C.\ Loizides is supported by the U.S. Department
of Energy, Office of Science, Office of Nuclear Physics, under
contract number DE-AC02-05CH11231. 

\bibliographystyle{utphys}
\bibliography{biblio}{}
\end{document}